\documentclass[12pt,reqno]{article}
\usepackage{amscd}
\usepackage{amsthm}
\usepackage{xcolor}
\usepackage[colorlinks=true,linkcolor=webgreen, filecolor=webbrown,citecolor=webgreen]{hyperref}
\definecolor{webgreen}{rgb}{0,.5,0}
\definecolor{webbrown}{rgb}{.6,0,0}
\usepackage{amsfonts,amsmath,amssymb}
\usepackage{float}
\usepackage{latexsym}
\usepackage{pstricks}

\newtheorem{theorem}{Theorem}

\begin{document}
\begin{center}
\vskip 1cm{\LARGE\bf The Set Partitions: Solution for the sharing secret keys}\footnote{This work was supported by L'IFORCE Laboratory of USTHB University.} \\ \ \\
Sadek BOUROUBI\footnote{Corresponding author}, Fella CHARCHALI  \\
University of Sciences and
Technology Houari Boumediene,\\Faculty of Mathematics,\\
P.Box 32 16111 El-Alia, Bab-Ezzouar, Algiers, Algeria\\
\href{mailto:sbouroubi@usthb.dz or bouroubis@yahoo.fr}{\tt
sbouroubi@usthb.dz or bouroubis@yahoo.fr}\\ 
\href{mailto:fellacharchali@ymail.com}{\tt
fellacharchali@ymail.com}
\\ \
 \\
Nesrine BENYAHIA TANI \\
Faculty of Economics and Management Sciences,\\2
Ahmed Waked Street, Dely Brahim, Algiers, Algeria\\
\href{mailto:benyahiatani@yahoo.fr}{\tt benyahiatani@yahoo.fr}
\end{center}

\begin{abstract}
Confidentiality was and will always remain a critical need
in the exchanges either between persons or the official parties.
Recently, cryptology has made a jump, from classical form to the
quantum one, we talk about quantum cryptography. This theory,
although is perfectly safe, there are still binding limits of
implementation. In this paper, we developed a new cryptographic protocol, called $BCB12$ protocol, which
will be used to provide random keys shared via a classical channel, using the set partitions.
Each key can be long enough that the plain text in question, in purpose, for instance,
to hide then to transmit the secret information using the Vernam
cipher.\vspace{0.1cm}\vspace{0.1cm}
\end{abstract}

\section{Introduction}

Issues such as confidentiality and integrity of information have
been solved by cryptography. The certificate that the Vernam cipher
is unconditionally secure, has transformed the problem to ensure the
confidentiality of information to a problem of distribution of the
secret key used in the encryption process between two parties. Until
the eighties, one way to distribute the secret key, apart from hand
to hand, was to use algorithms whose security is based on the
computational complexity. The generated keys by such algorithms are
reasonably secret but not unconditionally secret.\vspace{0.2cm}\\
\noindent In the early seventies, Stephen Wiesner wrote conjugate coding
\cite{weis}, describing the basis for a new concept that will be
known to the world in the early eighty by quantum cryptography.
Cryptography was attached to a quantum concept by the fact it relies
on photons to transmit secret information instead of bits. Security
is guaranteed not by mathematical theorems, but by the fundamental
laws of physics as the Heisenberg uncertainty principle which
asserts that certain
quantities cannot be measured simultaneously.\vspace{0.2cm}\\
\noindent Charles H. Bennett (who knew about Wiesner's idea) and Gilles
Brassard took the subject in $1984$ \cite{bras}, where they show up
to the world the first protocol of quantum key distribution whose
security is unconditional because confidentiality is based on
impossibilities imposed by the laws of physics \cite{bent}. This
protocol was implemented in $1989$ over a distance of $32\;cm$ by
calling efforts of F. Bessette, L. Salvail and J. Smolin, a full
description of the
prototype was published two years later \cite{bentB}.\vspace{0.2cm}\\
\noindent All Quantum Key Distribution protocols consist of two phases
\cite{cq}:\begin{enumerate}
 \item Initially one of the two parties sends to the other party "quantum"
signals then perform certain measurements.
 \item In a second time the
two parties engage in classical treatment of measurement results.
\end{enumerate}

\section{Concept of unconditional security - Vernam Cipher}
The Vernam cryptosystem, also known as the disposable mask or The
One Time Pad Cipher, provides perfect security, despite its
simplicity. In its classic form, it is nothing but a very long
random sequence of letters, written on pages bound together to form
a block. The sender uses each letter of the mask in turn to encrypt
exactly one plain text character. The Vernam Cipher text $C$ is a
function of both the message $M$ and the key $K$.\vspace{0.2cm}\\
\noindent The  Vernam cipher was invented in $1917$ by an engineer of
$AT\;$\&$\; T$, Gilbert S. Vernam \cite{reff}, who thought it would
become widely used for automatic encryption and decryption of telegraph messages. The vernam cipher is a
polyalphabetic substitution cipher belongs to secret key cryptosystems. The principle of the encryption
algorithm is that if a random key is added to a message, the bits of
the resulting string are also random and bear no information about
the message. If we use binary logic, unlike Vernam who worked with
an alphabet of $26$ letters, the encryption algorithm $E$ can be
written as:
\begin{center}
$E_{K}(M) = (M_{1} \oplus k_{1},M_{2} \oplus k_{2},\ldots,M_{n} \oplus k_{n})\ mod\ 2,$
\end{center}
where $M = (M_{1},M_{2},\ldots,M_{n})$ is the message to encrypt,
and $K = (k_{1},k_{2},\ldots,k_{n})$ is the key consisting of random
bits. The message and the key are added bitwise modulo $2$, i.e., the
exclusive-OR. Decryption process $D$ of cipher text $C$ is the same
as encryption, it is given by:
\begin{center}$M = D_{K}(C) = (C_{1} \oplus k_{1},C_{2} \oplus k_{2},\ldots,C_{n} \oplus k_{n})\ mod\ 2.$
\end{center}
Perfect security is ensured via the concept of entropy introduced by
Shannon in $1949$ \cite{shan}. Later, Vernam has been used in almost
all military concerns. Vernam fits very well to the definition of a
secret system \cite{stebila}, a fact confirmed by the following
theorem \cite{vaudenay}:
\begin{theorem}
The Vernam cipher is unconditionally secure for any distribution of
plain text.
\end{theorem}

\noindent But just like any other cryptosystem, it has significant drawbacks
which can cause its vulnerability such as the key must be as long as
the message to encrypt; the cryptosystem becomes vulnerable if the
same key is used more than once and the safest way to transport the
key is the diplomatic bag which requires users from the diplomatic
sector once.\vspace{0.2cm}\\
\noindent To remedy major drawbacks of this cipher, we propose here a new protocol,
called $BCB12$ "Bouroubi Charchali Benyahia $2012$", which is based on the set partitions concept.
\section{$BCB12$ Protocol}
The protocol $BCB12$ inspired from the quantum protocol $BB84$ "Bennett Brassard 1984" is based
on the set partitioning problem which is NP-hard. The expected objective from the protocol is to product and to distribute
a secret key via a classical channel, that will be used to ensure confidential communications between the participants by interchanging messages
encrypted by the Vernam cipher.\vspace{0.2cm}\\
\noindent First, the two parties involved in the protocol must share $\pi=\{A_{1},A_{2},\ldots,A_{k}\}$, a
partition of a set $[n]=\{1,2,\ldots,n\}$ into $k$-disjoint blocks ($n$ is assumed to
be large enough). Traditional protagonists who must  run an exchange
of information in cryptography are Alice and Bob. Both are  involved
in the sending and receiving secret messages and of course Eve, the
intruder who wants to spy on Alice and Bob.\vspace{0.2cm}\\
\noindent Suppose Alice wants to send a message $M$ to Bob, so steps to follow
are:\begin{enumerate}
      \item Alice calculates the number of characters $L_{M}$ of the message $M$ to be
      encrypted.
      \item Alice fixes the parameter $m$ such that $m = L_{M}\times S$, where $S$ is a positive integer called amplification parameter, which we explain the role later.
      \item Alice generates randomly a sequence of integers between
      $1$ and $n$ of size $m$, and for each integer in the sequence, she sets in a
list $T_{A}$ the index of the block to which this element belongs in the
partition $\pi$.
      \item Alice sends the parameter $m$ to Bob.
      \item Bob in turn generates a random sequence of integers
between $1$ and $n$ of size $m$, and for each integer in the
sequence, he sets in a list $T_{B}$ the index of the block to which this element
belongs in the partition $\pi$.
      \item Bob sends the list $T_{B}$ to Alice.
      \item Alice receives  Bob's list, and compares it with hers. If there is correspondence,
      i.e, for the same index, she locates the same block in both lists, she
puts a $"+"$, if not, she puts a $"-"$. Doing so, she creates a new
list, said $T$, whose elements are $"+"$ and $"-"$, then we have:
\[
T(i)=\left\{
\begin{array}{c}
+,\;\;if\;T_{A}(i)=T_{B}(i),\\\\
-,\;\;if\;T_{A}(i)\neq
T_{B}(i).
\end{array},\ \forall i=1,...,m.
\right.
\]
       \item Alice interprets each $"+"$ as the result of a function $f$ (defined below) chosen from
three functions (for example), acting on the elements of the
corresponding block. The concatenation of these results provides the random
secret key of length $L_{C}$.\vspace{0.2cm}\\
To identify $f$, Alice takes the first $"+"$ in the list $T$,
writes in binary the block index corresponding to this $"+"$, let be
$j$, then she considers the two first bit to the right.\vspace{0.2cm}\\
If the bits are:\vspace{0.2cm}\\
$i)$ identical ($"00"$ or $"11"$) then the function $f$ is
interpreted as the sum of elements of the block $A_{j}$.\\
$ii)$ $"10"$ then the function $f$ is interpreted as the product of the elements of the block $A_{j}$.\\
$iii)$ $"01"$ in this case, the function $f$ is interpreted as
the maximum element of the block $A_{j}$.
      \item Alice compares $L_{M}$ to $L_{C}$. If $L_{M}\leq L_{C}$, she sends
      the encrypted message and the sequence $T$ to Bob, and then, Bob performs step $8$ to get the same key.
       Otherwise, Alice must return to step $2$ and, at this level, she can keep the size $m$ or modify it.
    \end{enumerate}\vspace{0.2cm}

\noindent Absolute confidentiality requires a sharing of the key parameter
$\pi$ which ensures that the resulting key is secret
and is not known, only by legitimate users of the protocol.
Therefore, the generated key by $BCB12$ offers the privacy of
information transmitted, encrypted according to the Vernam
cryptosystem, ensuring inability to decrypt what was encrypted by a
spy. We Note finally that the generated keys have been proved random, using statistical tests. \\
\section{Illustrative example}
In this section, we present an illustrative and didactic example to show how $BCB12$ protocol runs, in order to get random secret keys, which will
be used for the plain texts ciphering. The first step, consists to generate a random partition of a set $[n]$
into $k$-blocks (for any $n$ and $k$ to choose). The second consists of unrolling the $BCB12$ protocol then inject the provided
key in Vernam, the adopted cryptosystem, to get out finally with the enciphered text.\\

\noindent Suppose now, Alice wants to send an encrypted message to Bob. We
consider first the following shared parameters between them: $n=20$, $k=13$ and the partition
$\pi=$\begin{scriptsize} \{\{1\};\ \{5\};\ \{14\};\  \{3\};\  \{10\};\ \{2\};\ \{6, 8, 12\};\  \{13, 18\};\
\{20\};\ \{9, 11\};\ \{15, 16, 19\};\  \{7\};\  \{4, 17\}\}\end{scriptsize}.\vspace{0.2cm}\\
\noindent Let be "\emph{\textbf{it rains take the umbrella}}" the secret message. The message written in binary form is:\vspace{0.2cm}\\
\begin{scriptsize}
01001001011101000010000001110010011000010110100101101110011100110010000001110100011000010110101101100101001000\\000111010001101000011001010010000001110101011011010110001001110010011001010110110001101100110000100101110,
\end{scriptsize}\vspace{0.2cm}\\
with length $L_{M}= 216$. Alice sets the parameter $S$ at $2$, it
follows that $m=216*2=432$.\vspace{0.2cm}\\ 
\noindent Alice generates her random sequence:\vspace{0.2cm}\\
\noindent \begin{scriptsize}\{12, 1, 4,
8, 10, 13, 16, 18, 18, 11, 13, 16, 15, 7, 4, 16, 8, 1, 13, 5, 17, 10,
2, 14, 7, 19, 11, 3, 16, 8, 1, 13, 6, 18, 10, 2, 15, 7, 11, 3, 15,
8, 20, 12, 4, 17, 9, 1, 14, 6, 18, 11, 3, 15, 6, 19, 11, 3, 16, 8, 4,
16, 9, 1, 13, 6, 18, 10, 2, 15, 7, 19, 2, 5, 17, 9, 2, 14, 6, 18,
12, 3, 16, 9, 20, 13, 5, 17, 9, 2, 14, 6, 19, 11, 3, 15, 7, 19, 12,
4, 16, 8, 19, 11, 4, 17, 9, 2, 14, 6, 18, 11, 3, 16, 8, 1, 13, 5,
17, 9, 2, 14, 6, 17, 9, 1, 13, 8, 20, 12, 4, 16, 8, 20, 12, 5, 17, 9,
1, 13, 5, 16, 8, 1, 13, 5, 17, 9, 2, 14, 6, 18, 10, 3, 15, 7, 18, 10,
2, 14, 7, 19, 11, 3, 15, 7, 19, 11, 3, 14, 6, 18, 9, 1, 13, 17, 11,
3, 2, 15, 8, 20, 16, 8, 20, 13, 5,17, 9, 2, 13, 6, 19, 12, 4, 16, 9,
1, 13, 6, 18, 13, 5, 17, 10, 2, 14, 6, 19, 11, 4, 16, 8, 1, 13, 5,
18, 10, 2, 15, 11, 3, 15, 8, 1, 13, 5, 18, 10, 2, 15, 7, 20, 12, 5,
17, 9, 2, 14, 7, 19, 12, 4, 16, 9, 1, 13, 6, 3, 15, 7, 20, 12, 4, 17,
9, 1, 16, 8, 20, 12, 5, 17, 9, 2, 14, 7, 19, 11, 4, 16, 9, 1, 14, 7,
19, 11, 3, 16, 7, 20, 12, 4, 16, 9, 2, 14, 7, 19, 12, 4, 16, 9, 1, 14,
10, 3, 15, 7, 20, 12, 4, 17, 9, 3, 16, 8, 20, 13, 5, 17, 9, 2, 14, 6,
18, 11, 3, 16, 8, 1, 14, 6, 19, 11, 4, 16, 8, 1, 13, 5, 18, 10, 2, 15,
7, 20, 12, 4, 17, 9, 2, 18, 10, 3, 15, 7, 20, 12, 4, 17, 11, 4, 16, 9,
1, 14, 6, 18, 10, 3, 15, 7, 20, 13, 5, 17, 10, 3, 15, 12, 4, 17, 9,
2, 14, 6, 19, 11, 3, 16, 8, 20, 13, 5, 18, 10, 2, 14, 7, 19, 11, 4, 16,
8, 20, 13, 5, 18, 11, 3, 16, 8, 20, 13, 5, 17, 9, 2, 14, 6, 18, 11,
3, 14, 8, 20, 12, 5, 17, 9, 1, 14, 6, 19, 11, 3, 18, 10, 2, 14\}.
\end{scriptsize}\vspace{0.2cm}\\
\noindent For each integer in the sequence, she sets in a list $T_{A}$ the
block index to which this element belongs in the partition
$\pi$.\vspace{0.2cm}\\
\noindent $T_{A}\ =\ $\begin{scriptsize}\{7, 1, 13, 7, 5, 8, 1, 7, 10, 8, 11, 11,
12, 13, 11, 7, 1, 8, 2, 13, 5, 6, 3, 12, 11, 10, 4, 11, 7, 1, 8, 7,
8, 5, 6, 11, 12, 10, 4, 11, 7, 9, 7, 13, 13, 10, 1, 3, 7, 8, 10, 4,
11, 7, 11, 10, 4, 11, 7, 13, 11, 10, 1, 8, 7, 8, 5, 6, 11, 12, 11,
7, 2, 13, 10, 6, 3, 7, 8, 7, 4, 11, 10, 9, 8, 2, 13, 10, 6, 3, 7,
11, 10, 4, 11, 12, 11, 7, 13, 11, 7, 11, 10, 13, 13, 10, 6, 3, 7, 8,
10, 4, 11, 7, 1, 8, 2, 13, 10, 6, 7, 13, 10, 1, 8, 7, 9, 7, 13, 11,
7, 9, 7, 2, 13, 10, 1, 8, 2, 11, 7, 1, 8, 2, 13, 10, 6, 3, 7, 8, 5,
4, 11, 12, 8, 5, 6, 3, 12, 11, 10, 4, 11, 12, 11, 10, 4, 3, 7, 8,
10, 1, 8, 13, 10, 4, 6, 11, 7, 9, 11, 7, 9, 8, 2, 13, 10, 6, 8, 7,
11, 7, 13, 11, 10, 1, 8, 7, 8, 8, 2, 13, 5, 6, 3, 7, 11, 10, 13, 11,
7, 1, 8, 2, 8, 5, 6, 11, 10, 4, 11, 7, 1, 8, 2, 8, 5, 6, 11, 12, 9,
7, 2, 13, 10, 6, 3, 12, 11, 7, 13, 11, 10, 1, 8, 7, 7, 11, 12, 9, 7,
13 , 13, 10, 1, 11, 7, 9, 7, 2, 13, 10, 6, 3, 12, 11, 10, 13, 11,
10, 1, 3, 12, 11, 10, 4, 11, 12, 9, 7, 13, 11, 10, 6, 3, 12, 11, 7,
13, 11, 10, 1, 3, 5, 4, 11, 12, 9, 9, 13, 13, 10, 4, 11, 7, 9, 8, 2,
13, 10, 6, 3, 7, 8, 10, 4, 11, 7, 7, 1, 3, 7, 11, 10, 13, 11, 7, 1,
8, 2, 8, 5, 6, 11, 12, 9, 7, 13, 13, 10, 6, 8, 5, 4, 11, 12, 9, 7,
13, 13, 10, 13, 11, 10, 1, 3, 7, 8, 5, 4, 11, 12, 9, 8, 2, 13, 5, 4,
11, 7, 13, 13, 10, 6, 3, 7, 11, 10, 7, 11, 7, 9, 8, 2, 8, 5, 6, 3,
12, 11, 10, 13, 11, 7, 9, 8, 2, 8, 10, 4, 11, 7, 9, 8, 2, 13, 10, 6,
3, 7, 8, 10, 4, 3, 7, 9, 7, 2, 13, 10, 1, 3, 7, 11, 10, 4, 8, 5, 6,
3\}.
\end{scriptsize}\vspace{0.2cm}\\
\noindent Alice sends $m$ to Bob. Bob in turn generates his random sequence of length $m$:\vspace{0.2cm}\\
\noindent \begin{scriptsize}\{6, 15, 20, 3, 6, 11, 13, 16, 18, 20, 8, 10, 14,
12, 3, 16, 8, 20, 11, 4, 15, 7, 19, 11, 3, 15, 6, 18, 10, 2, 14, 5,
17, 9, 1, 13, 4, 16, 8, 20, 12, 3, 15, 7, 19, 11, 2, 15, 7, 19, 11,
3, 15, 7, 18, 11, 4, 15, 8, 20, 13, 4, 16, 2, 14, 6, 19, 11, 3, 16,
8, 20, 12, 4, 17, 8, 20, 14, 9, 1, 13, 4, 17, 10, 2, 15, 7, 19, 11,
4, 16, 8, 1, 13, 5, 17, 10, 2, 14, 7, 19, 11, 3, 16, 6, 4, 17, 9, 1,
13, 6, 18, 10, 3, 15, 8, 1, 12, 5, 18, 10, 3, 15, 7, 20, 12, 4, 16,
9, 1, 13, 5, 18, 10, 2, 15, 7, 20, 12, 5, 17, 9, 2, 14, 6, 19, 11,
5, 17, 10, 3, 15, 7, 20, 12, 5, 17, 10, 1, 14, 7, 18, 11, 4, 19, 11,
4, 17, 3, 13, 7, 20, 12, 5, 18, 11, 3, 16, 9, 6, 18, 12, 19, 14, 7,
20, 13, 5, 8, 20, 13, 5, 17, 10, 2, 14, 6, 18, 11, 3, 15, 8, 20, 12,
4, 16, 8, 1, 13, 5, 18, 10, 3, 15, 7, 19, 12, 5, 17, 9, 1, 13, 6,
18, 10, 3, 15, 8, 20, 12, 4, 17, 9, 1, 14, 6, 18, 11, 3, 15, 7, 20,
12, 5, 17, 9, 1, 13, 5, 18, 9, 2, 14, 6, 18, 11, 2, 14, 6, 19, 11,3,
16, 8, 20, 12, 4, 17, 9, 1, 13, 5, 17, 10, 2, 15, 7, 19, 10, 3, 14,
7, 19, 11, 3, 15, 7, 20, 12, 4, 16, 9, 1, 13, 5, 17, 9, 1, 13, 5,
17, 9, 1, 13, 6, 17, 10, 2, 14, 6, 18, 11, 3, 15, 7, 20, 12, 4, 16,
8, 20, 13, 6, 18, 10, 3, 15, 7, 19, 11, 3, 15, 8, 19, 12, 4, 17, 10,
2, 15, 7, 19, 11, 3, 15, 7, 20, 11, 4, 17, 8, 1, 13, 5, 17, 9, 2,
14, 6, 18, 10, 2, 15, 7, 19, 10, 2, 14, 6, 19, 11, 3, 16, 8, 20, 12,
4, 16, 8, 1, 12, 4, 17, 10, 1, 15, 7, 18, 11, 3, 15, 7, 20, 12, 4,
16, 8, 20, 13, 5, 17, 9, 1, 13, 6, 17, 9, 1, 13, 5, 18, 10, 3, 15,
7, 19, 11, 3, 14, 7, 19, 12, 4, 17, 8, 1, 13, 5, 17, 10, 2, 14\}.
\end{scriptsize}\\

\noindent For each integer in the sequence, Bob sets in a list $T_{B}$ the
block index to which this element belongs in the partition $\pi$:\vspace{0.2cm}\\
\noindent $T_{B}\ =\ $\begin{scriptsize}\{7, 11, 9, 4, 7, 10, 8, 11, 8, 9, 7, 5,
3, 7, 4, 11, 7, 9, 10, 13, 11, 12, 11, 10, 4, 11, 7, 8, 5, 6, 3, 2,
13, 10, 1, 8, 13, 11, 7, 9, 7, 4, 11, 12, 11, 10, 6, 11, 12, 11, 10,
4, 11, 12, 8, 10, 13, 11, 7, 9, 8, 13, 11, 6, 3, 7, 11, 10, 4, 11,
7, 9, 7, 13, 13, 7, 9, 3, 10, 1, 8, 13, 13, 5, 6, 11, 12, 11, 10,
13, 11, 7, 1, 8, 2, 13, 5, 6, 3, 12, 11, 10, 4, 11, 7, 13, 13, 10,
1, 8, 7, 8, 5, 4, 11, 7, 1, 7, 2, 8, 5, 4, 11, 12, 9, 7, 13, 11, 10,
18, 2, 8, 5, 6, 11, 12, 9, 7, 2, 13, 10, 6, 3, 7, 11, 10, 2, 13, 5,
4, 11, 12, 9, 7, 2, 13, 5, 1, 3, 12, 8, 10, 13, 11, 10, 13, 13, 4,
8, 12, 9, 7, 2, 8, 10, 4, 11, 10, 7, 8, 7, 11, 3, 12, 9, 8, 2, 7, 9,
8, 2, 13, 5, 6, 3, 7, 8, 10, 4, 11, 7, 9, 7, 13, 11, 7, 1, 8, 2, 8,
5, 4, 11, 12, 11, 7, 2, 13, 10, 1, 8, 7, 8, 5, 4, 11, 7, 9, 7, 13,
13, 10, 1, 3, 7, 8, 10, 4, 11, 12, 9, 7, 2, 13, 10, 1, 8, 2, 8, 10,
6, 3, 7, 8, 10, 6, 3, 7, 11, 10, 4, 11, 7, 9, 7, 13, 13, 10, 1, 8,
2, 13, 5, 6, 11, 12, 11, 5, 4, 3, 12, 11, 10, 4, 11, 12, 9, 7, 13,
11, 10, 1, 8, 2, 13, 10, 1, 8, 2, 13, 10, 1, 8, 7, 13, 5, 6, 3, 7,
8, 10, 4, 11, 12, 9, 7, 13, 11, 7, 9, 8, 7, 8, 5, 4, 11, 12, 11, 10,
4, 11, 7, 11, 7, 13, 13, 5, 6, 11, 12, 11, 10, 4, 11, 12, 9, 10, 13,
13, 7, 1, 8, 2, 13, 10, 6, 3, 7, 8, 5, 6, 11, 12, 11, 5, 6, 3, 7,
11, 10, 4, 11, 7, 9, 7, 13, 11, 7, 1, 7, 13, 13, 5, 1, 11, 12, 8,
10, 4, 11, 12, 9, 7, 13, 11, 7, 9, 8, 2, 13, 10, 1, 8, 7, 13, 10, 1,
8, 2, 8, 5, 4, 11, 12, 11, 10, 4, 3, 12, 11, 7, 13, 13, 7, 1, 8, 2,
13, 5, 6, 3\}.
\end{scriptsize}\vspace{0.2cm}\\
\noindent The first integer obtained by Alice is $12$ which belongs to the block $A_{7}$, where the second one is $1$ which belongs to the block $A_{1}$. While,
the first integer obtained by Bob is $6$ belongs to the block $A_{7}$, and the second is $15$ belongs to the block $A_{11}$, and so on. Since we have the same index for the first integer in both sequences, Alice obtains the first $"+"$. By comparing the second integer obtained in both sequences, we can see that we have not the same index, so Alice put a $"-"$ in the second position. Doing so, Alice establishes the list $T$:\vspace{0.2cm}\\
\noindent $T\ =\ $
\begin{scriptsize}\{+, -, -, -, -, -, -, -, -, -, -, -, -, -, -,
-, -, -, -, -, -, -, -, -, -, -, -, -, -, -, -, -, -, -, -, -, -, +,
-, -, -, -, -, -, -, -, -, -, - , -, -, -, -, -, -, -, -, -, -, -,
-, +, +, -, -, -, -, -, -, -, -, -, -, -, -, -, -, -, -, -, +, -,
-, -, -, -, -, -, -, -, -, -, -, -, -, -, -, -, -, -, -, -, -, +, -,
+, +, +, -, -, +, +, -, +, +, +, +, -, +, -, -, -, -, -, -, -, -, -,
-, -, -, -, -, -, -, -, -, -, -, -, -, -, -, -, -, -, -, -, -, -, -,
-, -, -, -, -, -, -, -, -, -, -, -, -, -, -, -, -, -, -, -, -, -, -,
-, -, -, -, -, -, -, +, -, -, -, +, +, +, -, -, -, -, -, -, -, -, -,
-, -, -, -, -, -, -, +, -, -, -, -, -, -, -, -, -, -, -, -, -, -, -,
-, -, -, -, -, -, -, -, -, -, -, -, -, -, -, -, -, +, -, -, +, -, +,
-, -, +, +, +, -, -, -, -, -, +, -, -, -, -, -, -, -, -, -, -, -, -,
-, -, -, -, -, -, -, -, -, -, -, -, -, -, -, -, -, -, -, -, -, -, -,
-, -, -, -, -, -, -, -, -, -, -, -, -, -, -, -, -, -, -, -, -, -, -,
-, -, -, -, -, -, -, -, -, -, -, -, -, -, -, -, -, -, -, -, -, -, -,
-, -, -, -, -, -, -, -, -, -, -, -, -, -, -, -, -, -, -, -, +, -, +,
+, +, +, -, +, +, -, -, -, -, -, -, -, -, -, -, -, -, -, -, -,-, -,
-, -, -, -, -, -, -, -, -, -, -, -, -, -, -, -, -, -, -, -, -, -, -,
-, -, -, -, -, -, -, -, -, -, -, -, -, -, -, -, -, -, -, -, +, -, -,
-, -, +, +, +\}.
\end{scriptsize}\vspace{0.2cm}\\
\noindent As the first $"+"$ refers to the block $A_{7}$ and $7=00000111$ in
binary form, with the two first bit to the right are 11, then the function $f$ will be the sum of elements of blocks, hence:
\begin{equation*}
    f(A_{7})=6+8+12=26.
\end{equation*}
\noindent Therefore the provided key is:\vspace{0.2cm}\\
\noindent \begin{scriptsize}
\{26, 50, 21, 50, 31, 50, 21, 21, 20, 26, 31, 3, 50, 26, 1, 5, 26, 20, 31, 5, 21, 20,7, 26,
20, 1,31, 26, 20, 14, 26, 31, 10, 50, 7, 26, 10, 2, 14\}.
\end{scriptsize}\vspace{0.2cm}\\
\noindent Here the key is written in binary form, with length $L_{C}=312$:\vspace{0.2cm}\\
\noindent \begin{scriptsize}
000110100011001000010101001100100001111100110010000101010001010100010100000110100001111100
00001100110010000110\\100000000100000101000110100001010000011111000001010001010100010100000001110001101000010100
000000010001111100011\\0100001010000001110000110100001111100001010001100100000011100011010000010100000001000001110.
\end{scriptsize}\vspace{0.2cm}\\
\noindent Since $L_{C}>L_{M}$, Alice has to encrypt the
message. She gets the following encrypted text:\\

\noindent \begin{scriptsize}
010100110100011000110101010000000111111001011011011110110110011000110100011011100111111001101000 01010111001110\\1001110101011011010111111100110100011010100110100001110111011001100110001001110110011110000110000000110001.
\end{scriptsize}\vspace{0.2cm}\\
\noindent Then, she sends the list $T$ and the enciphered message to Bob.\\

\noindent Using the list $T$, Bob gets the key, therefore, he obtains the
plain text by performing the $Xor$
operation (operating principle of Vernam) between the enciphered message and the key.\\
\section{Conclusion}
Quantum cryptography ensures that the secret key is shared in
confidential way and an unauthorized party has not copy, the Vernam
Cipher, under restriction of eliminating its major drawbacks
mentioned above, offers unconditional security of the encrypted
message with assurance that without the possession the encryption
key, it is impossible to decipher what has been encrypted. The $BCB12$ protocol, carries out two objectives: The first objective being the production of a random key at least as long as the message
to be encrypted with assurance of the synchronization between the
transmitter and the receiver. So, the constraint mentioned above
will be removed. The second is the inability of a third person, said
Eve, to determine the secret key generated in a reasonable time.
This is because if Eve intercepts all data exchanged between Alice
and Bob, she has no information on the partition $\pi$ and the
secret key. To determine $\pi$ in order to find the key, Eve is
opposite to the following problem: Find all the $k$-blocks of a set
in the following  form $[k + i] = \{1, 2,\ldots, k + i\}\; i = 1 ,
2, ...,$ for each obtained partition, unroll the protocol in order
to generate all possible random keys and then, lead a exhaustive key
search, to find the right key, which is not feasible in a reasonable
time, at least during the lifetime of the shared secret between
Alice and Bob.

\end{document}